\documentclass[12pt,a4paper]{article}
 
\usepackage[utf8]{inputenc}
\usepackage[T1]{fontenc}
\usepackage{amsmath,amssymb,bm}
\usepackage[margin=1in]{geometry}
\usepackage{setspace}
\usepackage{hyperref}
\usepackage{graphicx}
\usepackage{booktabs}
\usepackage{caption}
\usepackage{xcolor}

\newcommand{\bu}{\bm{u}}
\newcommand{\bv}{\bm{v}}
\newcommand{\bn}{\bm{n}}
\newcommand{\bsig}{\bm{\sigma}}
\newcommand{\beps}{\bm{\varepsilon}}
\newcommand{\bI}{\bm{I}}
\newcommand{\bs}{\bm{s}}
\newcommand{\dev}{\mathrm{dev}}
\newcommand{\tr}{\mathrm{tr}}
 
\begin{document}
 
\title{Lightweight phase-field surrogate for modelling ductile-to-brittle transition\\
through phenomenological elastoplastic coupling}
 
\author{P G Kubendran Amos\footnote{Corresponding author: \texttt{prince@nitt.edu}} \\[4pt]
\small Theoretical Metallurgical Group, Department of Metallurgical and Materials Engineering, \\
\small National Institute of Technology Tiruchirappalli, Tamil Nadu, India 620015}
\date{}

\maketitle

\begin{abstract}
The ductile-to-brittle transition (DBT) in body-centred cubic systems is a central design constraint for cryogenic structures. 
Performing  parametric studies to enhance the understanding on DBT using fully coupled thermomechanical continuum DBT models is computationally expensive.
Therefore, in this work, a lightweight phase-field surrogate is proposed.
This surrogate approach captures key \emph{DBT-like} trends within a standard isothermal two-field (displacement--damage) setting by prescribing temperature dependence through three phenomenological mechanisms: (i) a temperature-dependent degradation exponent $n(T)$ that sharpens stiffness loss from gradual (ductile-like, $n=2.0$ at 293\,K) to abrupt (brittle-like, $n=3.5$ at 77\,K), (ii) temperature-dependent yield stress and elastic modulus to modulate the balance between plastic dissipation and elastic energy storage, and (iii) an effective fracture toughness and driving-force scaling to represent reduced crack-tip shielding at cryogenic temperatures. 
The model is implemented in FEniCSx using small-strain $J_2$ return mapping and a staggered solution scheme.
Simulations of a single-edge-notched specimen over 77--293\,K demonstrate a systematic progression from brittle-like to ductile-like response, characterised by reduced displacement to unstable fracture, a transition from abrupt post-peak load drop to extended softening, and a shift from narrow, localised damage bands with confined plasticity to broader process zones. 
A sensitivity study comparing four interpolation schemes (linear, smoothstep, exponential, hybrid) shows that the qualitative transition trends are robust, with interpolation primarily affecting intermediate-temperature responses while endpoint behaviours remain unchanged. 
By retaining only two coupled fields and local plasticity updates, the framework enables rapid temperature sweeps and design-space screening where fully coupled thermomechanical resolution is rather unnecessary or prohibitive.
\end{abstract}

\section{Introduction}
\label{sec:introduction}

\subsection{Cryogenic Applications}

The growing demand for technologies that operate under extreme cryogenic conditions has placed unprecedented requirements on the structural integrity of engineering materials. 
Nuclear fusion reactors such as the ITER rely on superconducting magnets cooled to temperatures as low as $-269\,^\circ$C~\cite{Iter2018,Devred2014}, while liquid hydrogen storage systems for clean energy applications expose containment vessels to temperatures near 20\,K~\cite{Mital2006}. 
The development of structural materials for such applications requires a thorough understanding of low-temperature failure mechanisms, particularly the ductile-to-brittle transition (DBT)~\cite{Reed1983,Ogata2014}.

As temperature decreases in body-centred cubic (BCC) systems, the thermally activated motion of dislocations is progressively suppressed, leading to a marked increase in yield strength and a corresponding reduction in the ability to exhibit plastic deformation~\cite{Cottrell1958,Rice1974}. 
The fracture mode consequently shifts from energy-absorbing ductile tearing to catastrophic brittle cleavage, in which cracks propagate rapidly with little prior warning~\cite{Knott1973,Anderson2005}. 

\subsection{Existing Works}

Extensive experimental investigations have characterised the DBT in structural alloys. While fracture toughness measurements using the master curve approach provide a statistically rigorous framework within the transition region~\cite{Wallin1999}, Charpy impact testing remains the most widely used method for determining the transition temperature~\cite{Wallin1984,Oldfield1975}. 
Studies on reduced-activation ferritic--martensitic steels for fusion applications have revealed that irradiation-induced hardening shifts the transition temperature upward~\cite{Tavassoli2004,Gaganidze2011}.
Despite this wealth of experimental data, computationally efficient screening tools that can capture the qualitative features of the DBT across a range of temperatures remain scarce.

Computational modelling of the DBT has advanced along several fronts. 
Molecular dynamics simulations have elucidated the temperature dependence of dislocation mobility in BCC metals~\cite{Guo2003}, but remain limited to nanometre-scale domains. 
Cohesive zone models~\cite{Needleman1990,Tvergaard1984}, crystal plasticity methods~\cite{Harewood2007}, and Gurson-type porous plasticity formulations~\cite{Beremin1983,Tanguy2005} have been employed at the continuum scale, but these approaches often require explicit crack tracking, element deletion, or involve computational costs that limit their applicability to large-scale simulations.

\subsection{Phase-field modelling}

The phase-field method for fracture, rooted in the variational theory~\cite{Francfort1998}, through appropriate implementation~\cite{Bourdin2000}, represents cracks as a diffuse damage field, eliminating the need for explicit crack tracking~\cite{Miehe2010a,Miehe2010b}. 
Recent investigations have systematically evaluated the predictive capabilities and limitations of phase-field fracture across a range of settings~\cite{Kristensen2023}. Extensions to elastoplastic fracture have introduced the coupling between plasticity and damage in varied forms including degradation of the elastic energy~\cite{Ambati2015ductile,Borden2016}, gradient-extended plasticity--damage theories at finite strains~\cite{Miehe2016,Aldakheel2018}, crystal-plasticity-coupled formulations for polycrystalline metals~\cite{Shanthraj2024}, and modifications to the crack driving force~\cite{You2021}.

With the incorporation of plasticity, several phase-field formulations capture elements of the brittle-to-ductile transition. 
A thermodynamically consistent, fully coupled thermo-elastoplastic phase-field technique has been developed, at finite strains, that modeled the failure mode transition under dynamic loading~\cite{Miehe2015}. 
This formulation solves a three-field problem (displacement, damage, temperature) with full heat conduction and thermomechanical coupling. 
A pressure-sensitive phase-field plasticity model has been proposed to simulate brittle-ductile transition largely for geomaterials~\cite{Choo2018}.
A recent technique introduces varying damage-driving energy coefficient to model this transition~\cite{You2021}.
These works are rigorous but share characteristics including multi-field coupling, pressure-dependent or finite-strain formulations,  that limit their direct application to large-scale structural simulations.
The present work attempts to address this need for a lightweight phase-field model that can reproduce the essential qualitative features of the temperature-driven DBT without the computational burden of fully coupled formulations.
Accordingly, a straightforward alternate in which temperature enters the standard isothermal elastoplastic phase-field framework solely through the calibration of a small number of temperature-dependent parameters is developed. 
More specifically, the DBT is modelled through (i)~a temperature-dependent degradation function $g(d,T) = (1-d)^{n(T)} + k_{\mathrm{res}}$; (ii)~temperature-dependent yield stress $\sigma_y(T)$ and elastic modulus $E(T)$; and (iii)~temperature-dependent effective fracture toughness $G_c^{\mathrm{eff}}(T)$ and driving force scaling $\alpha_\psi(T)$.

\subsection{Present Surrogate treatment}

Given the phenomenological coupling of thermomechanical driving-forces, the present deviates from the thermodynamic-consistency as claimed in the existing models~\cite{Miehe2015,Choo2018}. 
The desired thermomechanical coupling in this formulation is achieved through the  phenomenological knobs, like  temperature-dependent effective fracture toughness $G_c^{\mathrm{eff}}(T)$ and driving force scaling $\alpha_\psi(T)$.
These parameters, instead of being derived from a free-energy potential, are prescribed. 
Consequently, the energy functional, formulated in the surrogate approach, serves as a bookkeeping device for the staggered update rather than operating under true variational principle.
This surrogate treatment, in spite of \emph{not} explicitly resolving the temperature field, demonstrates that qualitative macroscopic signatures of the DBT can be captured within a standard two-field staggered framework by appropriate parameterisation.
This technique, therefore, can be adopted for generating preliminary design, parametric screening, and large-scale simulations wherein a complete and thorough thermomechanical resolution can be compromised.

The remainder of this paper is organised as follows: Sec.~\ref{sec:model} presents the model formulation; Sec.~\ref{sec:implementation} details the numerical implementation; Sec.~\ref{sec:validation} presents the verification studies; Sec.~\ref{sec:results} discusses the simulation results; Sec.~\ref{sec:temperature} presents the multi-temperature DBT analysis and interpolation sensitivity study; and Sec.~\ref{sec:conclusions} summarises the findings.

\section{Model Formulation}
\label{sec:model}

\subsection{Phase-field framework}

A two-dimensional body $\Omega \subset \mathbb{R}^2$ under plane strain conditions, containing a pre-existing notch, is considered. 
Following the AT2 regularisation~\cite{Ambrosio1990}, a sharp crack is approximated by a diffuse damage field $d(\bm{x}) \in [0,1]$, and the regularised crack surface functional is written as
\begin{equation}
\Gamma_\ell(d) = \int_\Omega \frac{1}{2\ell}\left( d^2 + \ell^2 |\nabla d|^2 \right) \mathrm{d}\Omega,
\label{eq:crack_surface}
\end{equation}
where $\ell$ is the regularisation length.

The model is solved in a staggered (alternating-minimisation) framework using a history-field treatment of irreversibility. 
Accordingly, the following \emph{effective} energy functional, which serves as a convenient bookkeeping device for the staggered update, is introduced
\begin{equation}
\mathcal{E}(\bu, d;T) =
\int_\Omega g(d,T)\,\psi^+(\beps^e)\,\mathrm{d}\Omega
+ \int_\Omega \psi^-(\beps^e)\,\mathrm{d}\Omega
+ \int_\Omega G_c^{\mathrm{eff}}(T)\,\frac{1}{2\ell}\!\left( d^2 + \ell^2 |\nabla d|^2 \right)\mathrm{d}\Omega,
\label{eq:total_energy}
\end{equation}
where $\psi^\pm$ are the tensile/compressive elastic energy contributions, $\beps^e = \beps - \beps^p$ is the elastic strain, and $g(d,T)$ is the temperature-dependent degradation function. 
As declared at the outset, temperature enters the model parametrically through the terms  like $G_c^{\mathrm{eff}}(T)$, instead of an explicitly solved field.

\subsection{Temperature-dependent degradation function}

In standard phase-field models, the degradation function takes the quadratic form $g(d) = (1-d)^2 + k_{\mathrm{res}}$~\cite{Miehe2010b}. 
For the intents of this work, a temperature-dependent generalisation is proposed, which reads
\begin{equation}
g(d, T) = (1 - d)^{n(T)} + k_{\mathrm{res}},
\label{eq:degradation}
\end{equation}
where the $n(T)$ is a temperature-dependent exponent. 
This exponent is referenced between 77 and 293 K and is written as
\begin{equation}
n(T) = n_{77} + (n_{293} - n_{77})\,\frac{T - 77}{293 - 77}, \quad T \in [77, 293]\,\text{K},
\label{eq:n_interp}
\end{equation}
wherein, the corresponding values  at bounding temperatures are $n_{293} = 2.0$ and $n_{77} = 3.5$.
This phenomenological exponent introduces the experimentally observed change from gradual post-peak softening (ductile response) to abrupt load loss (brittle response) at cryogenic temperatures. 
The derivative $\partial g/\partial d = -n(T)(1-d)^{n(T)-1}$ indicates that for $n>2$ the degradation remains close to unity for small $d$ but accelerates strongly as $d \to 1$, promoting snap-through-like post-peak behaviour.

\subsection{Temperature-dependent material properties}

Similar to exponent ($n(T)$) in above Eqn.~\ref{eq:n_interp}, the elastic and plastic properties are interpolated between reference values at 293\,K and 77\,K and are expressed as
\begin{align}
E(T) &= E_{77} + (E_{293} - E_{77})\,\frac{T - 77}{293 - 77}, \label{eq:E_T} \\
\sigma_y(T) &= \sigma_{y,77} + (\sigma_{y,293} - \sigma_{y,77})\,\frac{T - 77}{293 - 77}.
\label{eq:sy_T}
\end{align}
The reference values, representative of a low-carbon BCC steel, are listed in Table~\ref{tab:material}. 
The doubling of $\sigma_y$ from 350\,MPa to 700\,MPa reflects the thermally activated strengthening of BCC metals associated with the Peierls barrier~\cite{Cottrell1958,Reed1983}.

\begin{table}[htbp]
\centering
\caption{Temperature-dependent material parameters at reference temperatures.}
\label{tab:material}
\begin{tabular}{lcc}
\toprule
\textbf{Property} & \textbf{293\,K} & \textbf{77\,K} \\
\midrule
Young's modulus $E$ (GPa) & 210 & 220 \\
Yield stress $\sigma_y$ (MPa) & 350 & 700 \\
Poisson's ratio $\nu$ & 0.3 & 0.3 \\
Hardening modulus $H$ (MPa) & 1500 & 1500 \\
Degradation exponent $n$ & 2.0 & 3.5 \\
Effective toughness $G_c^{\mathrm{eff}}$ (N/mm) & 2.40 & 2.10 \\
Driving force scale $\alpha_\psi$ & 1.00 & 1.25 \\
\bottomrule
\end{tabular}
\end{table}

\subsection{Effective fracture toughness and driving force scaling}

The reduction in the plastic shielding at the crack tip with decrease in temperature is modelled by suitably formulating the effective fracture-toughness.
The temperature-dependent effective fracture-toughness is reads
\begin{equation}
G_c^{\mathrm{eff}}(T) = G_c \left( 0.9 + 0.3\,\frac{\sigma_{y,293}}{\sigma_y(T)} \right),
\label{eq:Gc_eff}
\end{equation}
where $G_c$ is the reference fracture energy. 
This formulation increases the apparent fracture resistance at higher temperature, where a larger plastic zone contributes to dissipation and crack-tip shielding.
Furthermore, the elastic driving-energy, that enters the history field, is scaled by
\begin{equation}
\alpha_\psi(T) = 1.0 + 0.5\left(1 - \frac{\sigma_{y,293}}{\sigma_y(T)}\right),
\label{eq:psi_scale}
\end{equation}
wherein at bounding temperatures, $\alpha_\psi(293\,\text{K}) = 1.0$ and $\alpha_\psi(77\,\text{K}) = 1.25$ is recovered. 
This scaling factor essentially provides a compact way to tune the onset of unstable crack growth under cryogenic conditions without introducing an additional thermal field.

\subsection{$J_2$ elastoplasticity}

The constitutive response of the undamaged material is described by small-strain $J_2$ plasticity with isotropic linear hardening. 
Cauchy stress tensor is accordingly expressed as
\begin{equation}
\bsig = 2\mu(T)\,\beps^e + \lambda(T)\,\tr(\beps^e)\,\bI,
\label{eq:elastic}
\end{equation}
with the yield function
\begin{equation}
f(\bsig,\bar{\varepsilon}^p)=\sigma_{\mathrm{eq}}-\left[\sigma_y(T)+H\bar{\varepsilon}^p\right]\le 0,
\end{equation}
where $\sigma_{\mathrm{eq}}=\sqrt{3\bs:\bs/2}$ is the von Mises equivalent stress and $\bs=\dev(\bsig)$ is the deviatoric stress. The associative flow rule is
\begin{equation}
\dot{\beps}^p=\dot{\gamma}\,\frac{3}{2}\frac{\bs}{\sigma_{\mathrm{eq}}},
\end{equation}
where $\dot{\gamma}\ge 0$ is the plastic multiplier, subject to the Kuhn--Tucker conditions that read $\dot{\gamma}\,f=0$ and $f\le 0$. Plastic variables are updated locally using a return-mapping algorithm.

\subsection{Energy decomposition and governing equations}

The elastic strain energy is decomposed into tensile and compressive contributions~\cite{Amor2009}.
Only the tensile part is allowed to drive damage, which prevents unphysical crack growth under compression. 
Using the elastic strain, $\beps^e=\beps-\beps^p$, the tensile driving energy is written as
\begin{equation}
\psi^+ = \tfrac{1}{2}K(T)\langle\tr(\beps^e)\rangle_+^2 + \mu(T)\,\dev(\beps^e):\dev(\beps^e),
\label{eq:psi_plus}
\end{equation}
where $K(T)$ and $\mu(T)$ are the bulk and shear moduli, $\dev(\cdot)$ denotes the deviatoric part, and $\langle\cdot\rangle_+=\max(\cdot,0)$ extracts the positive (tensile) volumetric contribution.

The coupled elastoplastic--phase-field problem is solved by a staggered (alternating) scheme.
At each load step, and for a fixed damage field $d$, including fixed plastic strain $\beps^p$ from the previous staggered iterate, the displacement field $\bu$ is obtained from the equilibrium equation
\begin{equation}
\nabla \cdot \left[ g(d,T)\,\bsig(\beps(\bu)-\beps^p) \right] = \bm{0},
\label{eq:momentum}
\end{equation}
where the degradation function $g(d,T)$ reduces the effective stiffness as damage grows.
Once $\bu$ is updated, the total strain $\beps(\bu)$ is known and the constitutive response is updated locally at quadrature points using return mapping, thereby yielding the updated stress $\bsig$, plastic strain $\beps^p$, and accumulated plastic strain $\bar{\varepsilon}^p$.

Damage irreversibility is enforced through a history variable, written as 
\begin{equation}
\mathcal{H}(\bm{x},t) = \max_{\tau \leq t}\{\alpha_\psi(T)\,\psi^+(\bm{x},\tau)\},
\end{equation}
stores the maximum tensile driving energy attained at each material point.
Correspondingly, once the material has experienced a given tensile driving force it cannot ``heal'' upon unloading. 
With this history field, the AT2 damage balance takes the form
\begin{equation}
G_c^{\mathrm{eff}}(T)\left(-\ell\Delta d + \frac{d}{\ell}\right) = 2\,\mathcal{H}\,(1-d),
\label{eq:damage_eq}
\end{equation}
where $G_c^{\mathrm{eff}}(T)$ controls the fracture resistance and $\ell$ sets the width of the regularised crack. 
In the staggered implementation, Eq.~\eqref{eq:damage_eq} is solved in its standard linearised history-field form~\cite{Miehe2010b}, while enforcing the bounds $d\in[0,1]$ and the irreversibility condition $d(t)\ge d(t-\Delta t)$.


\section{Numerical Implementation}
\label{sec:implementation}

\subsection{Geometry and discretisation}

The computational domain is a single-edge-notched specimen ($L \times H = 1.0 \times 0.2$\,mm) with a horizontal notch of length $a_0 = 0.20$\,mm and width $w = 0.02$\,mm at mid-height. 
All results assume plane strain conditions with unit out-of-plane thickness (1\,mm), so that reaction forces are reported in N and stresses in MPa (= N/mm$^2$).
The domain is discretised with unstructured linear triangular elements generated by Gmsh~\cite{Geuzaine2009} with characteristic size $h = 0.005$\,mm, yielding $\ell/h = 3.0$ for the regularisation length $\ell = 0.015$\,mm. 

The displacement field is approximated using first-order continuous Galerkin (CG1) vector elements, and the damage field using first-order CG1 scalar elements. 
The history variable is stored in a discontinuous, elementwise-constant DG0 space to promote robustness and suppress spurious oscillations. 
Internal plastic variables (plastic strain tensor $\beps^p$, equivalent plastic strain $\bar{\varepsilon}^p$, and Cauchy stress $\bsig$) are stored at quadrature points using Basix quadrature elements~\cite{Scroggs2022}. 
This choice enables stable local return-mapping updates without projection-induced noise.

\subsection{Initial damage seed}

A smooth radially decaying damage-seed, written as
\begin{equation}
d_0(\bm{x}) = 0.3\exp\!\left(-\frac{\|\bm{x} - \bm{x}_{\mathrm{tip}}\|}{0.5\ell}\right), \quad \|\bm{x} - \bm{x}_{\mathrm{tip}}\| < 1.5\ell,
\label{eq:seed}
\end{equation}
 is prescribed at the notch tip with $d_0 = 0$ elsewhere and $\max(d_0) = 0.30$. 
This damage seed triggers a preferred crack initiation site while maintaining a largely intact initial stiffness.

\subsection{Staggered solution scheme}

The coupled elastoplastic--phase-field problem is solved using an alternating-minimisation strategy with history-field irreversibility and under-relaxation~\cite{Miehe2010b}.
At each prescribed load increment and temperature $T$, we iterate between the displacement field and the damage field until both cease to change appreciably. 
Plasticity is treated locally by return mapping and is embedded into the staggered loop through a Picard-type eigenstrain coupling, so that plastic flow feeds back into mechanical equilibrium without introducing a fully nonlinear global solve.

Considering $(\bu^{(k)},d^{(k)},\beps^{p,(k)})$ denote the staggered iterates at iteration $k$ within a given load step, starting from the converged solution of the previous load increment, the following sequence is executed:

\textbf{(A) Displacement update with frozen damage and plastic strain.}
For a fixed damage field $d^{(k)}$ and plastic strain $\beps^{p,(k)}$, mechanical equilibrium is enforced in a linearised sense by treating $\beps^{p,(k)}$ as an eigenstrain. 
The displacement field $\bu^{(k+1)}$ is accordingly obtained from the weak form
\begin{equation}
\int_\Omega g(d^{(k)},T)\,\mathbb{C}(T):\beps(\bu^{(k+1)}):\beps(\bv)\,\mathrm{d}\Omega
=
\int_\Omega g(d^{(k)},T)\,\mathbb{C}(T):\beps^{p,(k)}:\beps(\bv)\,\mathrm{d}\Omega,
\label{eq:picard_u}
\end{equation}
which corresponds to $\nabla\cdot\!\left[g(d^{(k)},T)\,\bsig\!\left(\beps(\bu^{(k+1)})-\beps^{p,(k)}\right)\right]=\bm{0}$. 
This Picard-type coupling retains the essential physics of stress redistribution due to plastic flow while keeping the global displacement solve as a symmetric, linear problem.

\textbf{(B) Local constitutive update (return mapping).}
Given $\bu^{(k+1)}$, the total strain field $\beps(\bu^{(k+1)})$ is known.
The stress and internal variables are then updated pointwise at quadrature points by a standard $J_2$ return-mapping algorithm, producing the updated state $(\bsig^{(k+1)},\beps^{p,(k+1)},\bar{\varepsilon}^{p,(k+1)})$. 
This step captures yielding and hardening without any additional global unknowns.

\textbf{(C) History-field update for irreversibility.}
Using the updated elastic strain $\beps^e=\beps(\bu^{(k+1)})-\beps^{p,(k+1)}$, the tensile driving energy $\psi^+$ is evaluated and scaled by the temperature-dependent factor $\alpha_\psi(T)$ to define $\psi^{\mathrm{eff}}=\alpha_\psi(T)\psi^+$. Damage irreversibility is enforced by updating the history field pointwise as
\begin{equation}
\mathcal{H}^{(k+1)} \leftarrow \max\!\left(\mathcal{H}^{(k)},\,\psi^{\mathrm{eff}}\right),
\end{equation}
ensuring that previously attained tensile driving forces are retained upon unloading.

\textbf{(D) Damage update with frozen displacement and history.}
With $\bu^{(k+1)}$ and $\mathcal{H}^{(k+1)}$ fixed, the damage field is updated by solving the linearised AT2 history-field problem as
\begin{equation}
\int_\Omega G_c^{\mathrm{eff}}(T)\,\ell\,\nabla d^{(k+1)} \cdot \nabla \eta\,\mathrm{d}\Omega
+\int_\Omega \left(\frac{G_c^{\mathrm{eff}}(T)}{\ell}+2\mathcal{H}^{(k+1)}\right)d^{(k+1)}\,\eta\,\mathrm{d}\Omega
=
\int_\Omega 2\mathcal{H}^{(k+1)}\,\eta\,\mathrm{d}\Omega.
\label{eq:damage_weak}
\end{equation}
No Dirichlet conditions are imposed on $d$ with homogeneous Neumann conditions $\nabla d\cdot\bn=0$ are applied on $\partial\Omega$, allowing the crack phase-field to evolve freely up to the specimen boundaries.

\textbf{(E) Under-relaxation, bounds, and staggered convergence.}
To stabilise the staggered iterations, the damage update is under-relaxed,
\begin{equation}
d^{(k+1)} \leftarrow (1-\omega)\,d^{(k)} + \omega\,d^{\mathrm{trial}},
\end{equation}
with $\omega=0.4$. 
Irreversibility is then enforced strongly by $d^{(k+1)}\ge d_{\mathrm{old}}$ (where $d_{\mathrm{old}}$ is the converged damage from the previous load increment), and the admissible bounds $d^{(k+1)}\in[0,1]$ are imposed by clamping. 
The staggered loop is terminated once the relative changes in both fields satisfy the prescribed tolerance, i.e.\ $\max(\|\Delta\bu\|/\|\bu\|,\|\Delta d\|/\|d\|)<5\times10^{-5}$, or a maximum of 25 staggered iterations is reached.

The present strategy yields two low-cost linear solves per staggered iteration (one for $\bu$, one for $d$), while retaining elastoplastic dissipation through local return mapping and irreversibility through the history-field update.

Convergence is declared when $\max(\|\Delta\bu\|/\|\bu\|,\;\|\Delta d\|/\|d\|) < 5\times10^{-5}$, with a maximum of 25 staggered iterations per load step. Both sub-problems are solved with conjugate gradients (CG) preconditioned by HYPRE BoomerAMG~\cite{Falgout2002} at relative tolerance $10^{-10}$.

An early termination criterion halts the simulation when $R_y < 0.15R_y^{\mathrm{peak}}$ and $\max(d) > 0.95$ after at least 5 increments, preventing unnecessary post-failure iterations.

\subsection{Return-mapping algorithm}

At each quadrature point, plasticity is updated locally by a standard elastic predictor--plastic corrector (return-mapping) procedure.
Given the total strain at the current load step, $\beps$, and the previously converged internal variables $(\beps^p_n,\bar{\varepsilon}^p_n)$,  the increment is first assumed as purely elastic and form the \emph{trial} (elastic predictor) state by freezing the plastic strain at its old value. 
The corresponding elastic trial strain is
\[
\beps^{e,\mathrm{tr}}=\beps-\beps^p_n,
\]
and the trial stress follows from linear elasticity as
\begin{equation}
\bsig^{\mathrm{tr}} = 2\mu(T)\,\beps^{e,\mathrm{tr}}+\lambda(T)\,\tr(\beps^{e,\mathrm{tr}})\bI
=2\mu(T)\left(\beps-\beps^p_n\right)+\lambda(T)\tr\!\left(\beps-\beps^p_n\right)\bI.
\end{equation}
The trial stress is then checked against the yield condition using the trial equivalent stress $\sigma_{\mathrm{eq}}^{\mathrm{tr}}$ and the current hardening level as
\[
f^{\mathrm{tr}} = \sigma_{\mathrm{eq}}^{\mathrm{tr}}-\left[\sigma_y(T)+H\bar{\varepsilon}^p_n\right].
\]
If $f^{\mathrm{tr}}\le 0$, the point remains elastic and the trial state is accepted, i.e.\ $\bsig=\bsig^{\mathrm{tr}}$ and $(\beps^p,\bar{\varepsilon}^p)=(\beps^p_n,\bar{\varepsilon}^p_n)$.

If $f^{\mathrm{tr}}>0$, plastic loading occurs and the stress must be returned to the yield surface. 
Under associative $J_2$ plasticity with isotropic hardening, the return direction is radial in deviatoric stress space. 
The plastic consistency condition yields a closed-form increment of the plastic multiplier,
\begin{equation}
\Delta\gamma = \frac{\sigma_{\mathrm{eq}}^{\mathrm{tr}} - \sigma_y(T) - H\bar{\varepsilon}^p_n}{3\mu(T) + H},
\label{eq:return_map}
\end{equation}
and the deviatoric stress is corrected by moving back along the trial deviatoric direction.
Denoting $\bs^{\mathrm{tr}}=\dev(\bsig^{\mathrm{tr}})$, the stress is updated as
\begin{equation}
\bsig = \bsig^{\mathrm{tr}} - 3\mu(T)\Delta\gamma\,\frac{\bs^{\mathrm{tr}}}{\sigma_{\mathrm{eq}}^{\mathrm{tr}}}.
\end{equation}
The plastic strain tensor and accumulated plastic strain are then updated consistently with the associative flow rule,
\[
\beps^p = \beps^p_n + \Delta\gamma\,\frac{3}{2}\frac{\bs^{\mathrm{tr}}}{\sigma_{\mathrm{eq}}^{\mathrm{tr}}},
\qquad
\bar{\varepsilon}^p = \bar{\varepsilon}^p_n + \sqrt{\frac{2}{3}}\,\Delta\gamma,
\]
providing the local state $(\bsig,\beps^p,\bar{\varepsilon}^p)$ used in the subsequent history-field and damage updates.

Given the total strain $\beps$ and the previous plastic state $(\beps^p_n,\bar{\varepsilon}^p_n)$, the elastic trial stress is
\begin{equation}
\bsig^{\mathrm{tr}} = 2\mu(T)\left(\beps - \beps^p_n\right)+\lambda(T)\tr\!\left(\beps - \beps^p_n\right)\bI.
\end{equation}
If $f(\bsig^{\mathrm{tr}},\bar{\varepsilon}^p_n)\le 0$, the step is elastic. Otherwise, radial return gives
\begin{equation}
\Delta\gamma = \frac{\sigma_{\mathrm{eq}}^{\mathrm{tr}} - \sigma_y(T) - H\bar{\varepsilon}^p_n}{3\mu(T) + H},
\label{eq:return_map}
\end{equation}
with updated stress and corresponding updates to $\beps^p$ and $\bar{\varepsilon}^p$.


The present model is implemented in Python using FEniCSx/DOLFINx~\cite{Baratta2023}, UFL~\cite{Alnaes2014}, PETSc~\cite{Balay2019}, Gmsh~\cite{Geuzaine2009}, and Meshio~\cite{Schlomer2022}.


\section{Numerical Verification}
\label{sec:validation}

\begin{figure}[htbp]
\centering
\includegraphics[width=\textwidth]{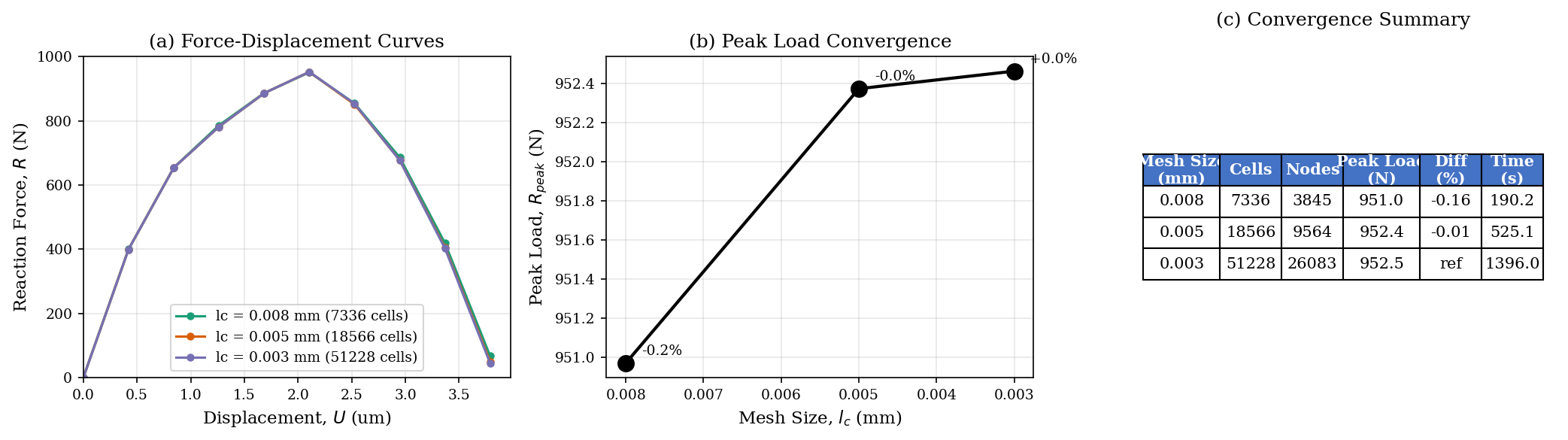}
\caption{Mesh convergence study at $T = 77$\,K. (a)~Force--displacement curves for three refinement levels. (b)~Peak load convergence. (c)~Summary table with wall-clock times.}
\label{fig:mesh_convergence}
\end{figure}

\subsection{Mesh convergence}

A convergence study is performed at $T = 77$\,K, sharpest nonlinearity in the present parameterisation, using three mesh levels (Table~\ref{tab:mesh}). 
This tabulation unravels a variation of less than 0.2\% in peak reaction across the three meshes indicating that the computed peak load is essentially mesh-independent for the adopted regularisation length and element size range.
However, it is vital to note that in order to keep the wall-clock time manageable on the finest mesh, this study employs a reduced displacement-controlled loading protocol of 20 increments up to $\bar{u}_{\max}=8$\,$\mu$m.
With a coarser increment size, the discrete sampling of the force--displacement response can under-resolve the exact peak location. 
Therefore, the peak loads reported in Table~\ref{tab:mesh} (approximately 952\,N) may differ slightly from those mentioned in the main results (Sec.~\ref{sec:results}), which use a finer load stepping. 

\begin{table}[htbp]
\centering
\caption{Mesh convergence results at $T = 77$\,K.}
\label{tab:mesh}
\begin{tabular}{lcccccc}
\toprule
$l_c$ (mm) & Cells & Nodes & $\ell/h$ & $R_{\mathrm{peak}}$ (N) & $\Delta R_{\mathrm{peak}}$ & Time (s) \\
\midrule
0.008 & 7\,336 & 3\,845 & 1.9 & 951.0 & $-0.16\%$ & 190 \\
0.005 & 18\,566 & 9\,564 & 3.0 & 952.4 & $-0.01\%$ & 525 \\
0.003 & 51\,228 & 26\,083 & 5.0 & 952.5 & ref & 1\,396 \\
\bottomrule
\end{tabular}
\end{table}

The force-displacements curves rendered by the difference meshes is plotted in Figure~\ref{fig:mesh_convergence}.
It is evident in this representation that the three force--displacement curves are virtually indistinguishable across the entire loading history, from the initial elastic regime through the peak load and into the post-peak softening branch.
All three meshes predict the same qualitative behaviour characterised by a linear elastic rise, a rounded peak near $\bar{u} \approx 2.0$\,$\mu$m, and a subsequent load drop as the damage field localises and propagates across the ligament.
Moreover, the close overlap in the post-peak softening trajectories indicate that the crack propagation speed and dissipated energy are consistent across the three discretisation.

The total variation in peak load across a nearly seven-fold increase in mesh density (from 7\,336 to 51\,228 cells) is less than 0.2\%. 
The convergence is monotonic from below, consistent with the expectation that coarser meshes slightly under-resolve the stress concentration at the crack tip, leading to marginally earlier damage initiation. 
The rapid convergence can be attributed to several factors.
Firstly, the AT2 model produces a smooth, exponentially decaying damage profile that is relatively benign to discretised treatments, compared to models with sharper crack representations such as the AT1 model~\cite{Ambati2015review}. 
The under-relaxed staggered scheme ($\omega = 0.4$) secondly ensures a gradual, stable evolution of the damage field, suppressing oscillatory artefacts that might otherwise arise from the sharp degradation-function at low temperatures. 
Thirdly, the volumetric-deviatoric energy decomposition~\cite{Amor2009} prevents spurious damage growth under compression, reducing the mesh sensitivity that can arise from incorrect energy splitting near the crack tip.

It is worth noting that the study was deliberately performed at 77\,K, where the elevated degradation exponent ($n = 3.5$) produces the sharpest stiffness degradation and the most severe nonlinearity in the coupled system. 
Convergence at this extreme temperature provides confidence that the model is at least equally well-converged at all intermediate and room-temperature conditions, where the smoother degradation places less stringent demands on the spatial resolution. 
The medium mesh ($l_c = 0.005$\,mm, $\ell/h = 3.0$) is adopted for all subsequent simulations, offering an optimal balance between accuracy and computational cost. 
The wall-clock times in Table~\ref{tab:mesh} confirm that the medium mesh completes in approximately 525\,s ($\approx$9\,minutes) on a single processor, making multi-temperature parametric sweeps practical.

\begin{figure}[htbp]
\centering
\includegraphics[width=\textwidth]{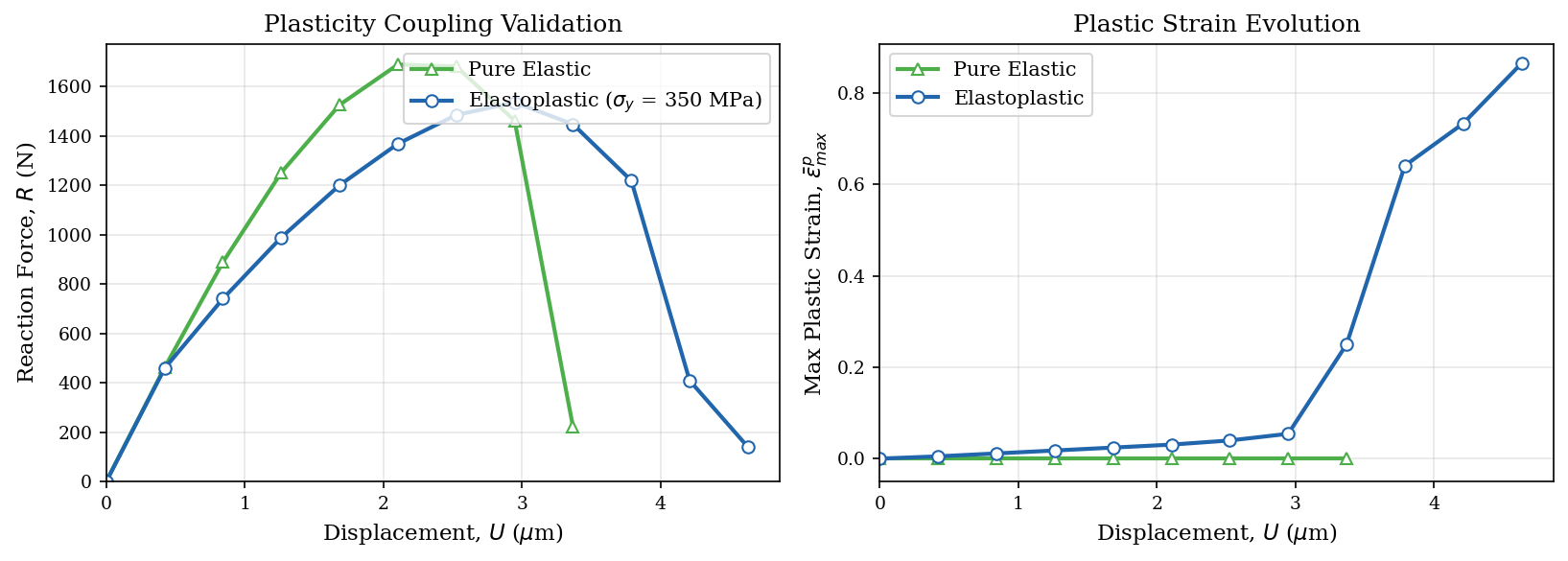}
\caption{Plasticity coupling verification at $T = 293$\,K. Left: force--displacement for pure elastic vs.\ elastoplastic models. Right: evolution of maximum equivalent plastic strain.}
\label{fig:plasticity_validation}
\end{figure}

\subsection{Plasticity coupling verification}

The accuracy of the elastoplastic coupling is assessed by comparing the $J_2$ model ($\sigma_y = 350$\,MPa) against a purely elastic simulation at 293\,K (Fig.~\ref{fig:plasticity_validation}). 
Both simulations employ identical geometry, mesh, phase-field parameters, and loading protocol.
The elastic case is recovered by setting the yield stress to an artificially high value that prevents plastic yielding at all load levels. 
This comparison isolates the effect of plasticity on the global fracture response.

Three physically consistent differences emerge from the comparison. 
Firstly, the pure elastic model reaches a peak load of approximately 1700\,N, whereas the elastoplastic model peaks at approximately 1490\,N, which is a reduction of roughly 12\%. 
This reduction arises because plastic yielding at the notch tip relaxes the local stress concentration, redistributing the load over a larger process zone and delaying the build-up of elastic strain energy that drives damage initiation. 
Secondly, the post-peak behaviour differs markedly with the elastic model exhibiting a sharp, abrupt load-drop, characteristic of brittle fracture, whereas the elastoplastic model displays a more gradual softening tail extending over a larger displacement range. 
This gradual softening is a hallmark of ductile fracture, where ongoing plastic dissipation in the process zone absorbs energy and retards crack propagation. 
Finally, the purely elastic case maintains exactly zero plastic strain throughout the entire loading history, confirming that no spurious yielding occurs when the yield criterion is not exceeded. 
In the elastoplastic case, the plastic strain remains negligible during the initial elastic phase, then accumulates gradually as the notch-tip stress reaches the yield surface, accelerating markedly in the post-peak regime to reach $\bar{\varepsilon}^p_{\max} \approx 0.9$. 
This acceleration reflects strain localisation within the fracture process zone, where as the effective stiffness degrades, the remaining intact material must accommodate an increasingly large fraction of the applied deformation.

These results confirm that the return-mapping algorithm is correctly integrated with the phase-field framework and that the elastoplastic coupling produces the expected qualitative modifications to the fracture response.

\begin{figure}[htbp]
\centering
\includegraphics[width=\textwidth]{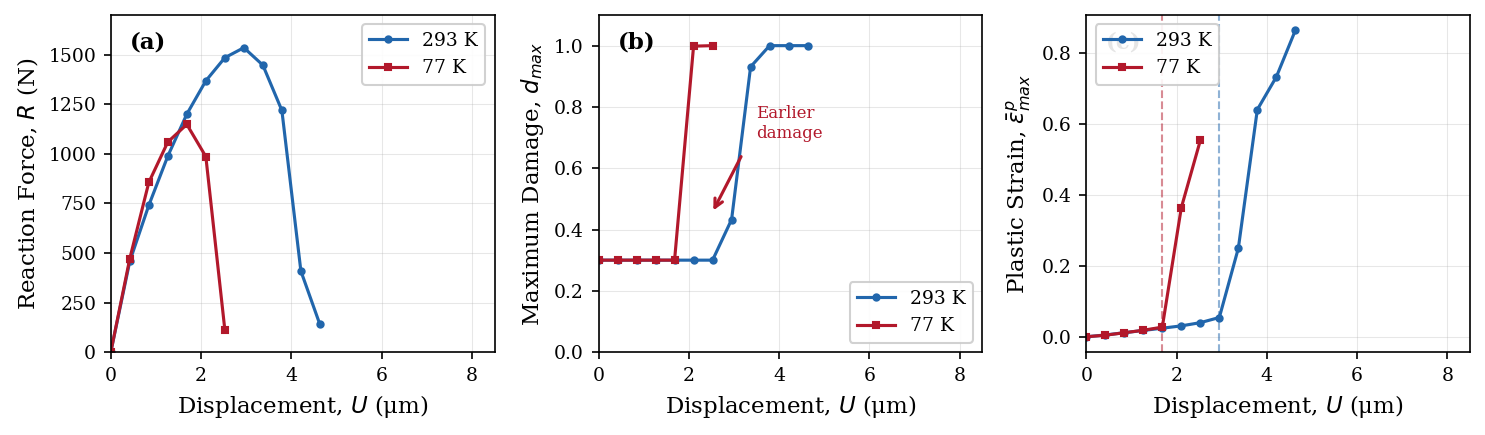}
\caption{Comparison of 293\,K and 77\,K simulations. (a)~Force--displacement curves. (b)~Evolution of maximum damage $d_{\max}$. (c)~Evolution of maximum equivalent plastic strain $\bar{\varepsilon}^p_{\max}$; dashed vertical lines mark the peak load for each temperature.}
\label{fig:three_panel}
\end{figure}


\section{Results: Ductile vs.\ Brittle Response}
\label{sec:results}

Model predictions at the two extreme temperatures, 293\,K (ductile-like) and 77\,K (brittle-like), is presented and discussed to demonstrate that the proposed lightweight framework captures the principal qualitative signatures associated with the temperature-driven ductile-to-brittle transition.

\subsection{Force--displacement response, damage, and plastic strain}

The macroscopic response of the present model at the distinct temperatures is shown in Figure~\ref{fig:three_panel} through force--displacement curves, evolution of maximum damage, and evolution of maximum equivalent plastic strain.

\textbf{Room temperature (293\,K).}
The force--displacement curve exhibits a nonlinear pre-peak regime as yielding develops ahead of the notch tip, reaching a peak of $\approx$1535\,N at $\bar{u} \approx 3.0$\,$\mu$m.
Post-peak softening extends over a substantial displacement range (up to $\bar{u}\approx 4.7$\,$\mu$m in the present simulation), indicating stable damage growth with significant dissipation.
The damage remains close to the seed value ($d_{\max}\approx 0.3$) up to $\bar{u}\approx 3.0$--$3.2$\,$\mu$m and then grows progressively, reaching $d_{\max}\approx 1.0$ at $\bar{u}\approx 3.8$--$4.0$\,$\mu$m.
The equivalent plastic strain increases gradually and accelerates in the post-peak regime, reaching $\bar{\varepsilon}^p_{\max}\approx 0.85$--$0.88$.

\textbf{Cryogenic temperature (77\,K).}
The initial stiffness is slightly higher ($E=220$\,GPa), and the peak load of $\approx$1150\,N is reached at a smaller displacement, $\bar{u}\approx 1.6$--$1.8$\,$\mu$m.
The subsequent load drop is abrupt, with the reaction force decreasing sharply over the next increment(s) and falling to $\mathcal{O}(10^2)$\,N by $\bar{u}\approx 2.5$\,$\mu$m.
The damage remains near the seed level ($d_{\max}\approx 0.3$) until $\bar{u}\approx 2.1$\,$\mu$m, where it jumps rapidly to $d_{\max}\approx 1.0$, indicating unstable crack advance with limited stable softening.
The equivalent plastic strain rises to $\bar{\varepsilon}^p_{\max}\approx 0.55$ by the final simulation step.
However, at peak load $\bar{\varepsilon}^p_{\max}$ remains below 0.1, and the subsequent rise reflects continued deformation of the localised process zone after the crack has fully traversed the ligament.

\textbf{Structural response vs.\ local material strength.}
A key observation is that the peak structural reaction force is lower at 77\,K despite the yield stress being doubled.
This reflects the distinction between a local material strength measure ($\sigma_y$) and a global structural quantity including peak reaction, which depends on the competition between load transfer through the ligament and crack propagation.
In the present model, the combined effect of reduced effective-toughness, amplified driving-force, and sharper stiffness-degradation promotes earlier unstable fracture at cryogenic temperature, limiting the extent to which the higher yield strength can be mobilised.
The von Mises stress field (Fig.~\ref{fig:vonmises}) confirms that the local stress scale at 77\,K approaches $\sim$700--800\,MPa, whereas the 293\,K case remains on a lower stress scale consistent with $\sigma_y=350$\,MPa and moderate hardening.

\subsection{Spatial evolution of the damage field}

\begin{figure}[htbp]
\centering
\includegraphics[width=\textwidth]{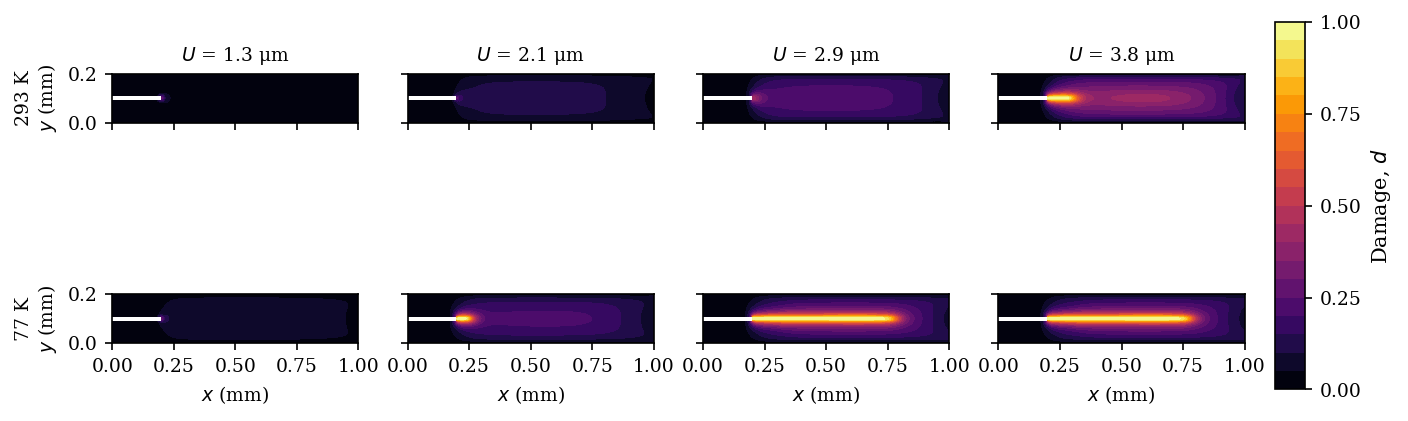}
\caption{Damage field $d(\bm{x})$ at four displacement levels ($\bar{u} = 1.3$, 2.1, 2.9, 3.8\,$\mu$m). Top row: 293\,K. Bottom row: 77\,K. Colour scale: $d = 0$ (dark, intact) to $d = 1$ (bright, fully fractured).}
\label{fig:damage_evolution}
\end{figure}

The progressive evolution of fracture/damage contours at four displacement levels under 293\,K and 77\,K is illustrated in Figure~\ref{fig:damage_evolution}.

At 293\,K, the damage field evolves gradually and exhibits a broad, diffuse process zone surrounding the advancing crack. 
At $\bar{u} = 1.3$\,$\mu$m, the damage is confined to the immediate vicinity of the notch tip, with the field barely exceeding its initial seed values. 
A diffuse damage zone has developed ahead of the notch by $\bar{u} = 2.1$\,$\mu$m, extending approximately one-third of the way across the ligament, with intermediate damage values ($d \approx 0.3$--$0.5$) spread over a wide region. 
By $\bar{u} = 2.9$\,$\mu$m, the damage band has propagated further and a narrow core of high damage ($d > 0.8$) is visible within the broader diffuse zone. 
The crack begins to traverse the ligament at $\bar{u} = 3.8$\,$\mu$m, but a substantial wake of partially damaged material ($d \approx 0.3$--$0.7$) persists around the crack path. 
This progressive, wide-zone damage morphology is consistent with a ductile-like response in the diffuse-crack setting: the large plastic zone around the crack tip dissipates energy and shields the crack from the applied driving force, resulting in stable, incremental crack growth where each step of propagation absorbs energy through both plastic work and fracture surface creation.

At 77\,K, the damage evolution follows a dramatically different pattern. 
At $\bar{u} = 1.3$\,$\mu$m, the damage field is still largely confined to the seed region, similar to the 293\,K case. 
However, by $\bar{u} = 2.1$\,$\mu$m, the damage has already localised into a narrow, intense band that extends well past the mid-point of the ligament, with the core reaching $d \approx 1.0$. 
The transition between the two snapshots is notably abrupt, consistent with the sharp load drop observed in the force--displacement curve over the same interval. 
By $\bar{u} = 2.9$ and $3.8$\,$\mu$m, the fully developed crack ($d = 1.0$) spans the entire ligament, with a much narrower damage zone compared to the 293\,K case. 
The surrounding material shows relatively little diffuse damage, indicating that the crack propagated rapidly through essentially undamaged material ahead of its tip rather than growing through a pre-softened plastic zone. 
This sharp, localised crack morphology is the spatial fingerprint of brittle-like fracture within the present phase-field representation~\cite{Anderson2005,Knott1973}.

The contrasting damage morphologies can be understood through the interplay of the three temperature-dependent mechanisms. 
At 293\,K, the low yield stress allows extensive plastic deformation ahead of the crack tip, which both absorbs energy and blunts the crack, while the gradual degradation function ($n = 2.0$) permits a broad process zone. 
At 77\,K, the elevated yield stress confines the plastic zone, the sharp degradation ($n = 3.5$) produces an abrupt stiffness drop once a critical damage level is reached, and the reduced effective toughness further lowers the energy barrier for crack extension.

\subsection{Von Mises stress distribution}

\begin{figure}[htbp]
\centering
\includegraphics[width=\textwidth]{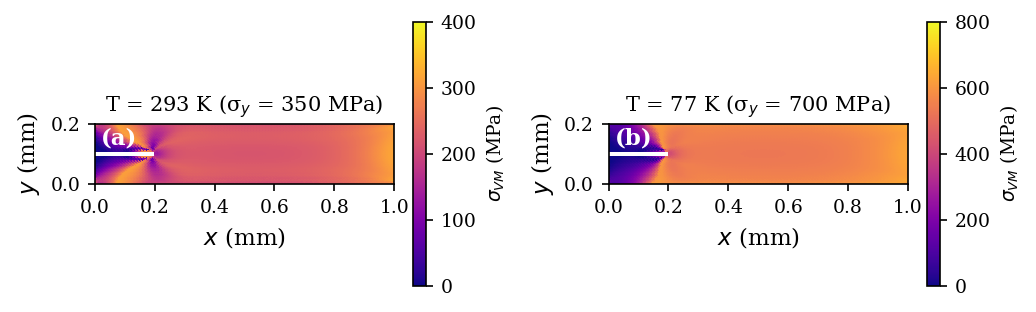}
\caption{Von Mises stress distribution at a representative pre-peak load step (cell-averaged). (a)~293\,K ($\sigma_y=350$\,MPa). (b)~77\,K ($\sigma_y=700$\,MPa). Note the different colour bar scales.}
\label{fig:vonmises}
\end{figure}

Figure~\ref{fig:vonmises} compares the von Mises stress at a representative pre-peak step for both temperatures. 
The colour scales differ between the panels to reflect the different yield stress levels.

At 293\,K (Fig.~\ref{fig:vonmises}(a)), the von Mises stress field exhibits a large yielded zone that extends well beyond the notch tip and spreads across a substantial fraction of the specimen cross-section.
The stress values in this region lie on a 0--400\,MPa scale, consistent with the yield stress $\sigma_y = 350$\,MPa and moderate isotropic hardening. 
The cell-averaged representation used in the plot provides a qualitative picture of the stress redistribution. 
The yielded zone is visible as the broad, relatively uniform warm-coloured region surrounding and extending ahead of the notch tip. 
More importantly, the stress field is diffuse rather than concentrated indicating that the plastic flow has redistributed the originally singular elastic stress concentration over a large volume of material. 
This extensive yielding serves as a natural crack-tip shielding mechanism~\cite{Rice1968}, absorbing a significant fraction of the externally applied energy through plastic work rather than channelling it into crack surface creation. 
The stress gradient ahead of the notch tip is mild, implying that the driving force for crack propagation builds up slowly, consistent with the gradual force--displacement response observed in Fig.~\ref{fig:three_panel}(a).

At 77\,K (Fig.~\ref{fig:vonmises}(b)), the stress field presents a strikingly different character.
The peak von Mises stress reaches approximately 700--800\,MPa, nearly double the room-temperature values, reflecting the elevated yield stress.
The high-stress region is crucially far more localised reflecting the plastic zone that is confined to a small volume in the immediate vicinity of the notch tip, while the bulk of the specimen remains at stress levels well below yield. 
The stress concentration is sharp and intense, with a steep gradient over a very short distance. 
This confinement is a direct consequence of the elevated yield stress, since a much higher stress is required to initiate yielding, and only a small volume of material near the geometric singularity reaches the yield surface. 
The remainder responds elastically, storing energy that is available for sudden release once the fracture criterion is met. 
The contrast between the diffuse stress field at 293\,K and the concentrated field at 77\,K mirrors the classical understanding of the DBT in BCC systems. 
At high temperatures, extensive plasticity blunts the crack tip and stabilises growth while confined plasticity preserves the stress concentration and enables unstable fracture at low temperatures~\cite{Cottrell1958}.

\section{Temperature Sweep and Interpolation Sensitivity}
\label{sec:temperature}

\subsection{DBT curve across four temperatures}

\begin{figure}[htbp]
\centering
\includegraphics[width=\textwidth]{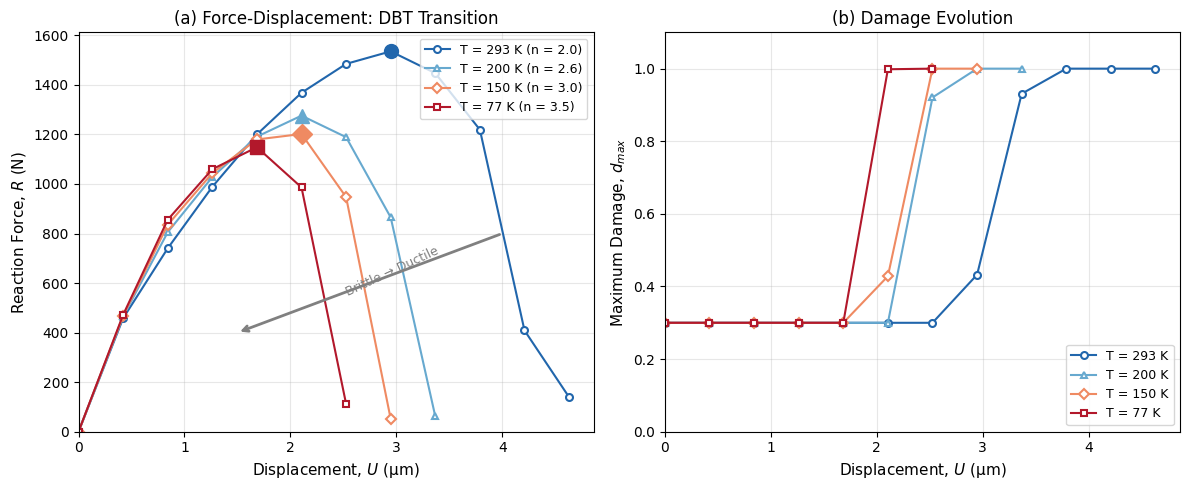}
\caption{(a)~Force--displacement curves across the temperature sweep. Arrow indicates the progression from brittle-like to ductile-like behaviour with increasing temperature. Filled symbols mark peak loads. (b)~Damage evolution $d_{\max}$ vs.\ displacement for all four temperatures.}
\label{fig:dbt_force_disp}
\end{figure}

To map the transition within the present model
The transition modelled by the present approach is further examined by simulating the damage evolution at four temperatures including 77, 150, 200, and 293\,K. Figure~\ref{fig:dbt_force_disp} presents the force--displacement curves and damage evolution for these four cases.

Across the sweep, the response progresses monotonically from brittle-like to ductile-like behaviour with increasing temperature. 
The peak load increases from 1148\,N at 77\,K through 1202\,N at 150\,K and 1276\,N at 200\,K to 1535\,N at 293\,K, representing a 34\% increase over the full temperature range.
The displacement at the onset of unstable fracture increases correspondingly, from approximately 1.6--2.1\,$\mu$m at 77\,K to beyond 3\,$\mu$m at 293\,K, indicating a progressive increase in ductility with rising temperature.

The intermediate temperatures reveal the transition mechanics most clearly. 
At 150\,K ($n = 3.0$, $\sigma_y = 581$\,MPa), the force--displacement curve occupies a position between the two extremes. 
The corresponding pre-peak stiffness and peak load are intermediate, and the post-peak load drop is rapid but not as abrupt as at 77\,K. 
A brief softening shoulder is visible before the load collapses, suggesting a narrow window of stable damage growth before the snap-through instability takes over. 
The damage evolution at 150\,K shows a steeper but still identifiable transition from the seed value to $d_{\max} = 1.0$, occurring over two to three load increments rather than the near-instantaneous jump observed at 77\,K. 
This intermediate behaviour corresponds to a mixed-mode failure regime in which the plastic zone is large enough to provide some energy absorption but too small to fully stabilise crack growth.

At 200\,K ($n = 2.6$, $\sigma_y = 500$\,MPa), the response shifts further toward the ductile end. 
The peak load of 1276\,N is reached at a larger displacement, and the post-peak softening extends over several load steps with a discernible gradual character. 
The damage evolution shows a progressive, multi-step growth pattern that is qualitatively similar to the 293\,K case, though compressed into a shorter displacement range. 
The force--displacement curve at 200\,K closely resembles that at 293\,K in shape but is shifted to lower load levels and smaller displacements, indicating that the transition from brittle-like to ductile-like behaviour is largely complete by this temperature for the present parameter set.

The damage evolution curves in Fig.~\ref{fig:dbt_force_disp}(b) provide a complementary perspective. 
At 77\,K, the damage remains at the seed value until a critical displacement is reached, at which point it jumps to $d_{\max} = 1.0$ within one or two steps. 
At 150\,K, the jump is still present but preceded by a short incubation period of gradual growth. 
At 200\,K and 293\,K, the damage grows progressively over many steps, with the transition from seed to full fracture spanning a displacement range of approximately 1--2\,$\mu$m. 
This systematic change in the damage evolution rate is a direct consequence of the degradation exponent. 
Particularly, higher $n$ values produce a sharper threshold between the stable and unstable branches of the damage response, compressing the transition into fewer load increments.

\begin{figure}[htbp]
\centering
\includegraphics[width=\textwidth]{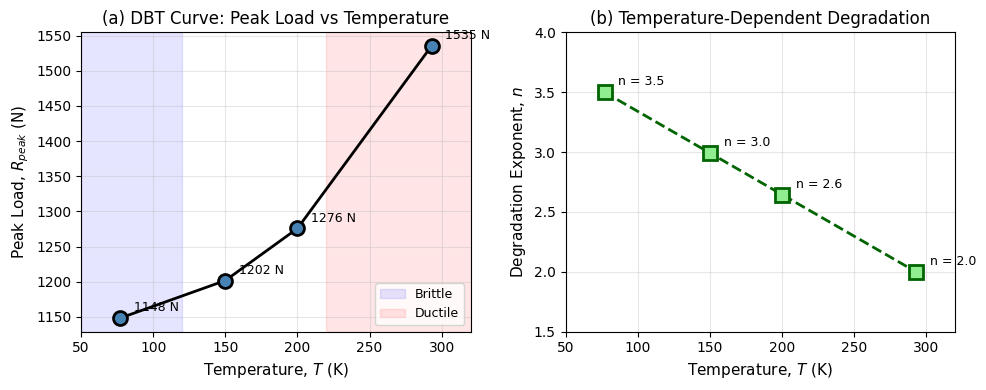}
\caption{(a)~Peak load vs.\ temperature for the present specimen and loading protocol. (b)~Temperature-dependent degradation exponent $n(T)$.}
\label{fig:dbt_curve}
\end{figure}

Figure~\ref{fig:dbt_curve} presents the peak load vs.\ temperature curve alongside the degradation exponent schedule.

The peak load--temperature curve exhibits a sigmoidal-like progression between lower and upper regimes within the present model. 
The curve rises steeply between 77\,K and 200\,K, where the peak load increases by approximately 128\,N (from 1148 to 1276\,N), and then continues to rise more gradually toward 293\,K, gaining a further 259\,N. 
The inflection region lies near 150--200\,K, which is consistent with the typical DBT range reported for low-carbon structural steels~\cite{Anderson2005}. 
The qualitative progression from a low-temperature brittle-like regime to a high-temperature ductile-like regime mirrors the general structure of DBT curves reported for ferritic steels. 

The degradation exponent schedule in Figure~\ref{fig:dbt_curve}(b) illustrates the underlying parameterisation.
The exponent $n$ decreases linearly from 3.5 at 77\,K to 2.0 at 293\,K, and smoothly interpolating between the brittle-like and ductile-like degradation regimes. 
The near-linear relationship between $n$ and $T$ in the default scheme produces the sigmoidal peak-load curve because the structural response is nonlinearly sensitive to the degradation exponent. 
The transition from stable to unstable damage evolution occurs over a relatively narrow range of $n$ values near $n \approx 2.5$--$3.0$.

\begin{figure}[htbp]
\centering
\includegraphics[width=\textwidth]{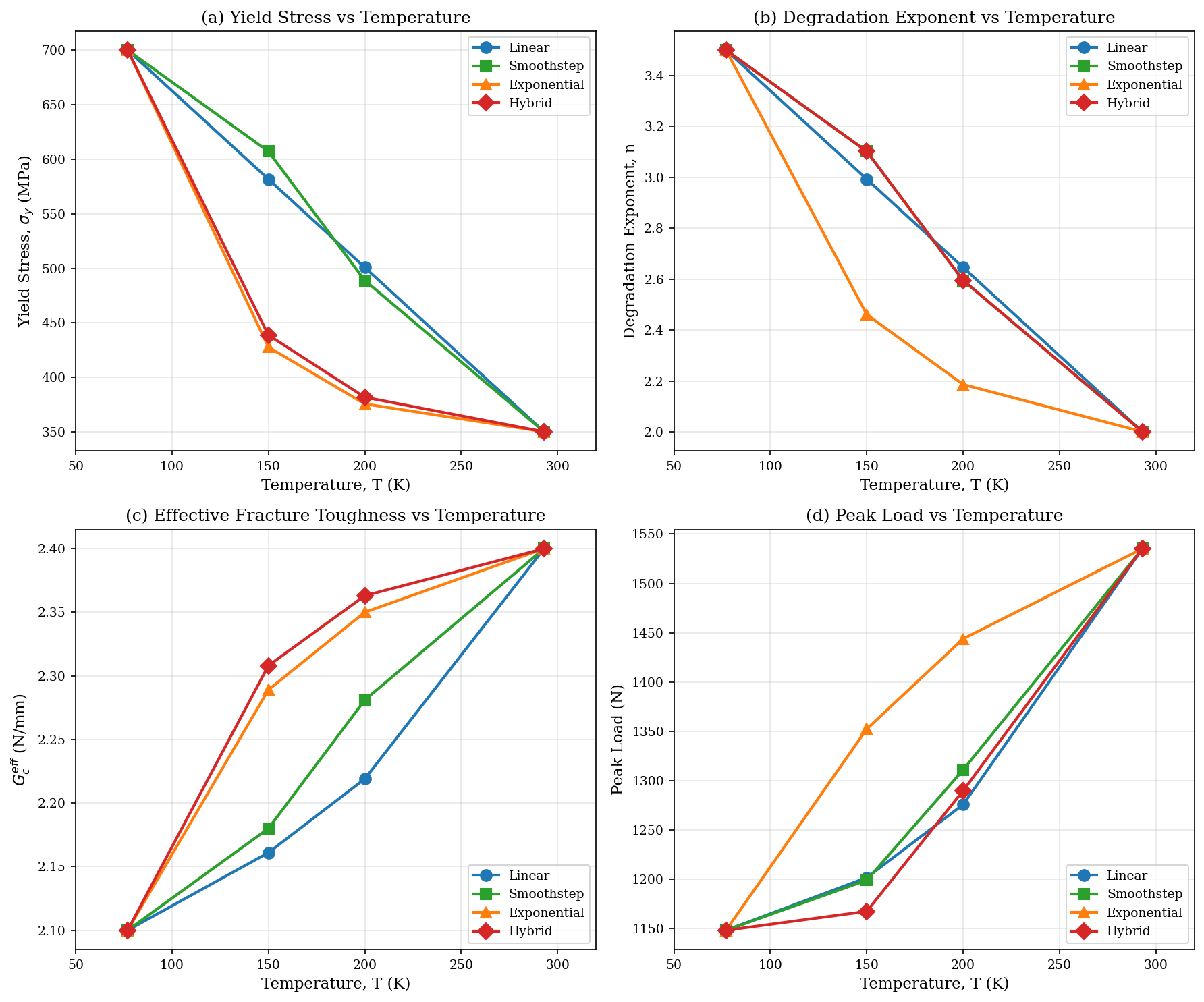}
\caption{Effect of interpolation scheme on the temperature dependence of: (a)~yield stress, (b)~degradation exponent, (c)~effective fracture toughness, and (d)~predicted peak load.}
\label{fig:interp_params}
\end{figure}

\subsection{Sensitivity to interpolation scheme}
\label{sec:interpolation}

The temperature-dependent parameters in the model ($\sigma_y(T)$, $n(T)$, $G_c^{\mathrm{eff}}(T)$, $\alpha_\psi(T)$) are interpolated between their reference values at 77\,K and 293\,K. 
To quantify the influence of this modelling choice at intermediate temperatures, four interpolation schemes (linear, smoothstep, exponential, hybrid), all of which reproduce identical endpoint values by construction, are compared.

\begin{figure}[htbp]
\centering
\includegraphics[width=\textwidth]{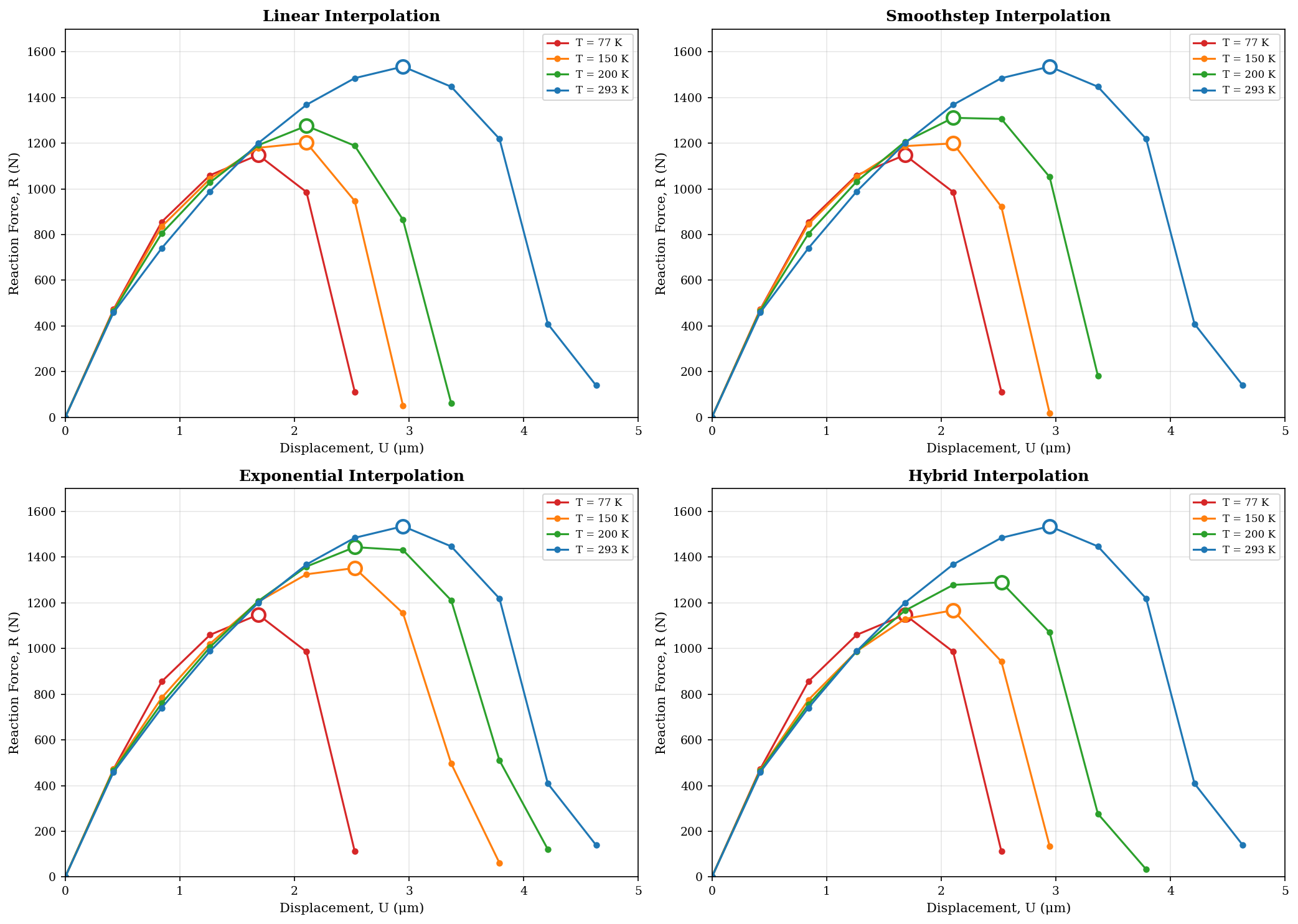}
\caption{Force--displacement curves at four temperatures for each interpolation scheme: linear (top-left), smoothstep (top-right), exponential (bottom-left), and hybrid (bottom-right). Open circles mark peak loads.}
\label{fig:interp_force}
\end{figure}

Figure~\ref{fig:interp_params} shows the parameter schedules and the resulting peak load--temperature curves.
The corresponding force--displacement curves are presented in Figure~\ref{fig:interp_force}.

Three conclusions follow from this sensitivity analysis. 
Firstly, the 77\,K and 293\,K responses coincide across all schemes by construction, confirming interpolation-independence at the endpoints. 
This is expected since all four schemes reproduce identical parameter values at the boundary temperatures.
Nevertheless, the similar curves serve as a coherent check for the internal consistency of the formulation.

The intermediate-temperature responses secondly shift quantitatively depending on the interpolation choice. 
The exponential scheme moves more rapidly toward the 293\,K parameter values and therefore produces higher peak loads at 150--200\,K than the linear scheme (e.g., approximately 1350\,N vs.\ 1200\,N at 150\,K), with smoothstep and hybrid lying between. 
This spread of approximately 150\,N at 150\,K represents the interpolation-induced uncertainty in the model predictions at intermediate temperatures. 
The spread narrows at 200\,K (to approximately 100\,N) and vanishes at the endpoints, indicating that the interpolation sensitivity is concentrated in the steepest part of the transition.

Thirdly, and most importantly, the qualitative brittle-to-ductile progression is preserved in all four cases. 
Every scheme produces a monotonic increase in peak load with temperature, a progressive transition from abrupt to gradual post-peak softening, and a consistent identification of the transition region near 150--200\,K. 
This robustness indicates that the model's DBT predictions are governed primarily by the endpoint constitutive parameters and the three-mechanism coupling structure rather than by the specific functional form of the interpolation. 
In practice, the interpolation choice would be informed by experimental calibration data for a given alloy.
The linear scheme is adopted as the default throughout this work for its simplicity and transparency.

\section{Discussion}
\label{sec:discussion}

\subsection{Qualitative fidelity}

The present phenomenological formulation reproduces four commonly cited macroscopic signatures of the ductile-to-brittle transition:

(i) \emph{Peak load reduction at low temperature.} Despite the doubled yield stress at 77\,K, the peak structural reaction force is lower than at 293\,K (1148\,N vs.\ 1535\,N).
In the present specimen and parameterisation, this occurs because unstable fracture intervenes before the full plastic load-carrying capacity can be mobilised.

(ii) \emph{Reduced ductility.} The displacement at fracture decreases from $>$4\,$\mu$m at 293\,K to approximately 1.6--2.1\,$\mu$m at 77\,K, representing a reduction of more than 50\% in the overall ductility.

(iii) \emph{Transition in post-peak behaviour.} The load drop is gradual and extended at 293\,K (stable ductile-like tearing) and abrupt at 77\,K (unstable brittle-like cleavage). The intermediate temperatures (150\,K, 200\,K) exhibit progressively more gradual softening with increasing temperature, mapping out the transition region.

(iv) \emph{Change in fracture morphology.} The damage zone is broad and diffuse at 293\,K (large process zone with partially damaged material surrounding the crack path) and narrow and localised at 77\,K (confined crack band with minimal surrounding damage).

These qualitative features are consistent with the experimental phenomenology of the DBT in ferritic steels as characterised by Charpy impact tests~\cite{Wallin1984,Anderson2005} and fractographic observations that reveal a shift from dimpled (ductile) to transgranular cleavage (brittle) fracture surfaces with decreasing temperature~\cite{Knott1973}. 
The peak load--temperature trend exhibits a sigmoidal-like progression between the brittle-like and ductile-like regimes for the adopted parameter set.

\subsection{Role of the individual mechanisms}

The model achieves the DBT through the concerted action of three phenomenological mechanisms, each contributing a distinct aspect of the transition.

The \emph{degradation exponent} $n(T)$ primarily controls the character of the post-peak response. 
At $n \approx 2.0$ (room temperature), the standard quadratic degradation produces a smooth, progressive loss of stiffness with increasing damage, allowing stable damage evolution over many load increments. 
The sharpened degradation, at $n \approx 3.5$ (cryogenic temperature), creates a threshold-like behaviour wherein the material retains most of its stiffness until damage reaches a critical level.
Subsequently, a rapid cascade of stiffness loss and damage growth occurs, producing an abrupt post-peak load drop characteristic of brittle fracture. 
The sensitivity analysis in Sec.~\ref{sec:interpolation} indicates that the critical range of $n$ where the transition from stable to unstable behaviour occurs lies near $n \approx 2.5$--$3.0$, corresponding to temperatures of approximately 150--200\,K in the linear interpolation scheme.

The \emph{yield stress} $\sigma_y(T)$ governs the extent of the plastic zone and, consequently, the degree of crack-tip shielding. 
The doubling of $\sigma_y$ from 350 to 700\,MPa between 293\,K and 77\,K reduces the plastic zone size by approximately a factor of four, since the plastic zone radius scales as $r_p \sim (K_{IC}/\sigma_y)^2$~\cite{Irwin1960}.
This change fundamentally alters the balance between plastic dissipation and fracture energy release and is directly visible in the von Mises stress maps (Fig.~\ref{fig:vonmises}), wherein the extensive, diffuse yielded region at 293\,K that redistributes stress and shields the crack tip contracts to a small, confined zone at 77\,K that preserves the stress concentration.

The \emph{effective toughness} $G_c^{\mathrm{eff}}(T)$ and \emph{driving force scaling} $\alpha_\psi(T)$ provide the quantitative calibration that positions the transition correctly in the displacement space. 
The reduced toughness at 77\,K (2.10 vs.\ 2.40\,N/mm) lowers the energy barrier for crack propagation, while the elevated driving force ($\alpha_\psi = 1.25$) ensures that the critical threshold is reached earlier in the loading history.
Without these two parameters, the model would still produce qualitative differences between ductile-like and brittle-like behaviour through $n(T)$ and $\sigma_y(T)$ alone, but the quantitative placement of the peak load and the displacement at fracture would be less controllable.

\subsection{Comparison with existing formulations}

One of the key distinction from fully coupled thermo-elastoplastic phase-field formulations, and the motivation for the present surrogate, is computational. 
Rigorous thermomechanical models resolve additional coupled fields (e.g.\ temperature) and nonlinear couplings, thereby substantially increasing per-simulation cost. 
The present framework intentionally trades thermodynamic rigour for speed, while retaining only two global unknown fields ($\bu$ and $d$) with local return-mapping plasticity updates, making it suitable for rapid parametric screening and preliminary design studies. 
Table~\ref{tab:comparison} summarises the principal differences.

\begin{table}[htbp]
\centering
\caption{Comparison with existing phase-field formulations for failure mode transitions.}
\label{tab:comparison}
\begin{tabular}{p{2.8cm}p{2.4cm}p{2.4cm}p{2.4cm}p{2.4cm}}
\toprule
\textbf{Feature} & \textbf{Miehe et al.\ \cite{Miehe2015}} & \textbf{Choo \& Sun \cite{Choo2018}} & \textbf{You et al.\ \cite{You2021}} & \textbf{Present work} \\
\midrule
Coupled fields & 3 ($\bu$,$d$,$T$) & 2 ($\bu$,$d$) & 2 ($\bu$,$d$) & 2 ($\bu$,$d$) \\
Strain formulation & Finite & Small & Small & Small \\
Plasticity model & $J_2$ + thermal & Drucker--Prager & $J_2$ & $J_2$ \\
Heat conduction & Yes & No & No & No \\
DBT driver & Temperature field & Confining pressure & Energy coefficient & $T$-dep.\ $n$, $\sigma_y$, $G_c$ \\
Thermodynamic consistency & Full & Full & Partial & Phenomenological \\
\bottomrule
\end{tabular}
\end{table}

\subsection{Computational efficiency}

A central focus of the proposed approach is computational simplicity. 
Each temperature simulation involves only two coupled fields (displacement and damage) solved in a staggered fashion, with the plasticity handled locally at quadrature points via a standard return-mapping algorithm.
No thermal field equation is solved, no finite-strain geometric nonlinearity is involved, and the staggered sub-problems are linear systems amenable to efficient CG/AMG solvers.

The wall-clock times reported in Table~\ref{tab:mesh} provide a concrete measure of the computational cost. 
On the adopted medium mesh (18\,566 cells, 9\,564 nodes), a complete simulation at a single temperature requires approximately 525\,s ($\approx$9\,minutes) on a single processor. 
A full four-temperature sweep spanning the DBT range is therefore completed in under 40\,minutes. Even the finest mesh (51\,228 cells) completes in approximately 23\,minutes, remaining well within practical limits. 
By contrast, the fully coupled thermo-elastoplastic formulation of Miehe et al.~\cite{Miehe2015} solves three coupled fields with finite-strain kinematics and transient heat conduction, substantially increasing both the per-step cost and the implementation complexity. 
While a direct timing comparison is not available (different codes, hardware, and problem sizes), the reduction from three coupled nonlinear fields to two coupled linear sub-problems with local plasticity represents a qualitative reduction in computational complexity that enables the kind of parametric sweeps and sensitivity analyses presented in Sec.~\ref{sec:temperature}.

\subsection{Resulting Trade-off}

The phenomenological approach that offers computational efficiency introduces noticeable trade-off.

Firstly, while the functional forms are motivated by the underlying physics (Peierls barrier strengthening, plastic zone shielding), the temperature-dependent parameters ($n(T)$, $G_c^{\mathrm{eff}}(T)$, $\alpha_\psi(T)$) are phenomenologically prescribed rather than derived from thermodynamic principles. 
Correspondingly, their specific numerical values require calibration against experimental data for a given material. 
In the present study, representative values for low-carbon BCC steels are used.
However, a rigorous application to a specific alloy (e.g., EUROFER97 for fusion applications) would require calibration against Charpy or fracture toughness data across the transition temperature range.

The model secondly operates under the assumption of isothermal conditions at each temperature. 
Transient thermal effects, most notably adiabatic heating during rapid plastic deformation near the crack tip, are not captured. 
In real materials, this adiabatic heating can locally raise the temperature and partially counteract the cryogenic embrittlement, an effect that would require the solution of a coupled thermal field equation to resolve.

Third, the small-strain assumption may become inaccurate in the immediate vicinity of the crack tip where the equivalent plastic strain reaches values of $\bar{\varepsilon}^p \approx 0.5$--$0.9$. 
At such strain levels, finite deformation effects (geometric nonlinearity, stress--strain measure conjugacy) can influence the local stress state and the crack-tip driving force. The present model accepts this approximation as adequate for the qualitative DBT characterisation that is its intended purpose.

Finally, the model does not incorporate explicit microstructural features --- grain size, crystallographic texture, precipitate distributions, or grain boundary characteristics --- that are known to influence the DBT temperature in real alloys. 
In particular, the well-documented effect of grain refinement in suppressing the DBT~\cite{Anderson2005} cannot be captured within the present continuum framework without additional constitutive enrichment.

These limitations define clear directions for future enrichment, while the present formulation serves its intended role as a lightweight surrogate for qualitative DBT screening and parametric studies.

\section{Conclusions}
\label{sec:conclusions}

A lightweight phase-field model has been developed for the ductile-to-brittle transition in BCC steels through three phenomenological mechanisms involving a temperature-dependent degradation exponent, temperature-dependent yield stress and elastic properties, and temperature-dependent effective fracture toughness with driving force scaling. 
The model requires only two coupled fields (displacement and damage) within a staggered isothermal framework.

In spite of its phenomenological approach, the model captures key qualitative DBT-like trends: reduced peak load despite increased material strength, diminished ductility, transition from gradual softening to abrupt post-peak load drop, and shift from diffuse to localised damage morphology.
Furthermore, the peak load--temperature curve exhibits a sigmoidal-like progression qualitatively reminiscent of experimental DBT curves, with a transition region near 150--200\,K for the adopted material parameters.

A sensitivity study on four interpolation schemes demonstrates that the predicted DBT behaviour is robust to the choice of interpolation, with all schemes producing qualitatively consistent transition curves that differ only at intermediate temperatures.

The structural paradox of lower peak load at higher yield stress is explained through the distinction between local material strength and global structural load-carrying capacity wherein at cryogenic temperatures, fracture intervenes before the full plastic capacity can be mobilised.

Future work will focus on calibrating the phenomenological parameters against experimental Charpy and fracture toughness data for specific alloys, extending the model to three dimensions, incorporating rate-dependent effects for dynamic loading, and benchmarking against the fully coupled thermo-elastoplastic formulation of Miehe et al.~\cite{Miehe2015} for quantitative validation.

\section*{Data availability}

The simulation code and data supporting this study are available from the corresponding author upon reasonable request.

\section*{Acknowledgement}
The author thanks the Anusandhan National Research Foundation (ANRF) for its financial support through the Advanced Research Grant. 

\section*{Funding}
This work was supported by ANRF under ANRF/ARG/2025/000213/ENS.

\section*{Declaration of competing interest}
The authors declare that they have no known competing financial interests or personal relationships that could have appeared to influence the work reported in this paper.

\appendix

\section{Summary of model parameters}
\label{app:parameters}

Table~\ref{tab:all_params} provides a complete summary of all numerical parameters used in the simulations.

\begin{table}[htbp]
\centering
\caption{Complete list of simulation parameters.}
\label{tab:all_params}
\begin{tabular}{lll}
\toprule
\textbf{Parameter} & \textbf{Symbol} & \textbf{Value} \\
\midrule
\multicolumn{3}{l}{\textit{Geometry}} \\
Specimen length & $L$ & 1.0\,mm \\
Specimen height & $H$ & 0.2\,mm \\
Notch length & $a_0$ & 0.20\,mm \\
Notch width & $w$ & 0.02\,mm \\
\midrule
\multicolumn{3}{l}{\textit{Phase-field}} \\
Regularisation length & $\ell$ & 0.015\,mm \\
Reference fracture energy & $G_c$ & 2.0\,N/mm \\
Residual stiffness & $k_{\mathrm{res}}$ & $10^{-7}$ \\
Degradation exponent (293\,K) & $n_{293}$ & 2.0 \\
Degradation exponent (77\,K) & $n_{77}$ & 3.5 \\
\midrule
\multicolumn{3}{l}{\textit{Elastoplasticity}} \\
Young's modulus (293\,K) & $E_{293}$ & 210\,GPa \\
Young's modulus (77\,K) & $E_{77}$ & 220\,GPa \\
Yield stress (293\,K) & $\sigma_{y,293}$ & 350\,MPa \\
Yield stress (77\,K) & $\sigma_{y,77}$ & 700\,MPa \\
Poisson's ratio & $\nu$ & 0.3 \\
Hardening modulus & $H$ & 1500\,MPa \\
\midrule
\multicolumn{3}{l}{\textit{Numerical}} \\
Element size & $h$ & 0.005\,mm \\
Staggered tolerance & $\mathrm{TOL}_{\mathrm{alt}}$ & $5 \times 10^{-5}$ \\
Max staggered iterations & -- & 25 \\
Under-relaxation & $\omega$ & 0.4 \\
Solver tolerance & -- & $10^{-10}$ \\
Damage seed maximum & $d_0^{\max}$ & 0.30 \\
Seed radius & $r_{\mathrm{seed}}$ & $1.5\ell$ \\
\bottomrule
\end{tabular}
\end{table}

\section{Interpolation scheme definitions}
\label{app:interpolation}

Let $\xi = (T - 77)/216 \in [0,1]$ be the normalised temperature, and let $p_{77}$ and $p_{293}$ denote the endpoint values of any parameter $p$. The four interpolation schemes used in Sec.~\ref{sec:interpolation} are defined as:

\begin{align}
\text{Linear:} \quad p(T) &= p_{77} + (p_{293} - p_{77})\,\xi, \\
\text{Smoothstep:} \quad p(T) &= p_{77} + (p_{293} - p_{77})\,(3\xi^2 - 2\xi^3), \\
\text{Exponential:} \quad p(T) &= p_{77} + (p_{293} - p_{77})\,\frac{1 - e^{-3\xi}}{1 - e^{-3}}, \\
\text{Hybrid:} \quad &\begin{cases} \text{Exponential for } \sigma_y, G_c^{\mathrm{eff}}, \\ \text{Smoothstep for } n, \alpha_\psi. \end{cases}
\end{align}

All schemes satisfy $p(77) = p_{77}$ and $p(293) = p_{293}$ by construction.


\end{document}